\documentclass[prl,twocolumn]{revtex4}

\usepackage{amsmath}
\usepackage{amssymb}
\usepackage{graphicx}

\begin{document}
\author{Thomas Salzburger and Helmut Ritsch}

\title{An atom-photon pair laser}

\affiliation{Technikerstrasse 25/2, 6020 Innsbruck, Austria}
\date{\today}
\begin{abstract}
We study the quantum dynamics of an ultracold atomic gas in a deep optical lattice within an optical high-$Q$ resonator.
The atoms are coherently illuminated with the cavity resonance tuned to a blue vibrational sideband, so that photon scattering to the resonator mode is accompanied by vibrational cooling of the atoms.
This system exhibits a threshold above which pairwise stimulated generation of a cavity photon and an atom in the lowest vibrational band dominates spontaneous scattering and we find a combination of optical lasing with a buildup of a macroscopic population in the lowest lattice band.
Including output coupling of ground-state atoms and replenishing of hot atoms into the cavity volume leads to a coherent, quantum correlated atom-photon pair source very analogous to twin light beam generation in a nondegenerate optical parametric oscillator.
\end{abstract}

\maketitle
Soon after the experimental realization of Bose-Einstein condensation of ultracold atoms, a coherent atomic beam was produced by continuous RF out-coupling of atoms from a BEC in a magnetic trap \cite{BECoutcoupled}.
Although duration of such quasi continuous atom lasing is limited by the total number of condensed atoms it constituted the first realization of an atom laser, soon followed by closely related implementations \cite{atomlaserfollowups,Ottl}. Even earlier in theory atom lasers attracted a great deal of interest and a variety of proposals to implement continuous atom lasers were made \cite{atomlasertheory}. In practice a central hurdle towards a cw implementation is the presence of near resonant light in the cooling process, which hampers the buildup of a degenerate matter wave state by reabsorption heating if densities sufficient for evaporation are involved. One way to avoid this is spatial separation of the laser and evaporative cooling stages in a long beam geometry. This should allow continuous operation \cite{ENS} but the requirements to reach sufficient atomic flux for fast evaporation in the final stage are challenging.   

In this Letter we study a scheme involving concurrent generation of a coherent atomic beam and a laser beam in a single device. It consists of laser cooled atoms in a far off-resonant optical lattice with some spatial region of the lattice confined within an optical cavity. Lasing on an anti-Stokes vibrational Raman transition should dominate spontaneous emission and replace evaporation in the last cooling stage to reach atomic quantum degeneracy \cite{Horak,Salzburger2}. Thus the cavity solves the reabsorption problem and lowers density requirements as no collisional thermalization is needed. 

In principle it has been noted already some time ago that ground state cooling of a single trapped particle can be enhanced by employing a cavity \cite{Cirac}.
This scheme, proposed to operate in the bad cavity regime, has been recently generalized and analyzed in more detail also for narrow bandwidth cavities \cite{Zippilli,Murr06}.
In both cases the cavity acts as a passive element. As low cavity photon numbers are required this cannot be simply scaled to larger volumes with many particles.
Nevertheless in a transient regime collective enhancement of cavity side band cooling was predicted in a low excitation regime \cite{Beige}.
In a parallel development we studied light forces in a single-atom laser involving an incoherently pumped pointlike atom in a cavity. When optical gain is tied to motional cooling,  lasing, cooling and trapping occur together \cite{Salzburger} with final temperatures below the standard cavity cooling limit \cite{Salzburger2}.

In this work we combine these ideas and study a Raman laser operated on the vibrational anti-Stokes transition of the quantized trapped atom states of an optical lattice.
This combines gain enhanced cavity cooling with collective amplification of ground-state occupation and suppressed spontaneous emission. Above threshold
we expect buildup of a coherent cavity field and a macroscopic atomic population utilizing Bose enhancement from both atoms and photons. 

Note that for large detuning of the Raman transition from the upper atomic state, cavity loss as the main dissipation channel acts like a diode on the vibrational transition.
Cavity photons and ground-state atoms are produced in pairs but photons have a smaller lifetime than atoms in the lowest band, so that a much larger atom population will build up.
Atomic out-coupling is required to prevent self termination of the laser. In addition atomic heating will enhance the photon gain and accelerate cooling. 

\begin{figure}
  \includegraphics[width=8cm]{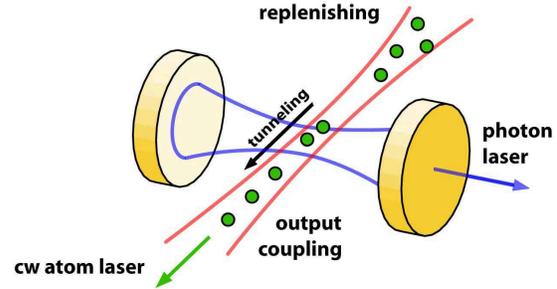}
  \caption{Schematic of the atom-photon pair laser. Hot atoms tunneling into the resonator accumulate in the lowest (laser) band while a coherent cavity field builds up.}
  \label{fig:zero}
\end{figure}

\begin{figure}
  \includegraphics[width=8cm]{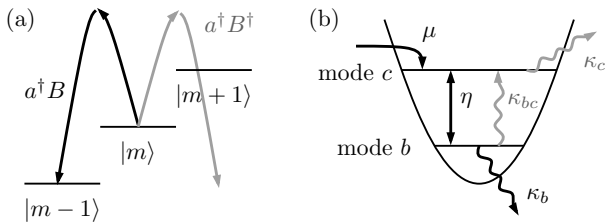}
  \caption{(a) Transition scheme for the Raman gain.
    When the cavity-laser frequency difference matches the energy gap, heating (gray path) is suppressed.
    (b) Illustration of the atomic dynamics.}
  \label{fig:one}
\end{figure}

Let us now construct a simple model Hamiltonian.
First we have the two lasing modes for the photons and atoms, i.e.\ a single resonator mode for the light and the lowest lattice band for the atom laser mode.
Some pre-cooling assures sufficiently many atoms in the lowest two vibrational states of the lattice, which at this point is a 1D lattice of atoms transverse to the cavity mode (see Fig.~\ref{fig:zero}) as discussed as well in Refs.~\cite{selforg,seesaw}.
In the simplest case the lattice laser also takes the role of the Raman pump and thus is assumed to be red detuned from the cavity mode by the energy difference between the two atomic bands.
Raman gain induces motion-photon coupling so that in the Lamb-Dicke limit \cite{javanainen1981lct} each scattering event induces a concurrent change of the atomic motional state and photon number by one.
An illustration of the Raman gain transition can be found in Fig.~\ref{fig:one} (a) where $m$ indicates the motional quantum number of the atom.

A corresponding effective single-particle Hamiltonian for a  far detuned Raman transition coupling the cavity mode and the motional atomic dynamics reads \cite{reviewarticle}
\begin{gather}
  H_0=\frac{p^2}{2m}+V_0\sin^2(kx)+\hbar(\Delta_c+U_0)a^\dagger a\nonumber\\
  +i\sqrt{\hbar V_0U_0}\,\sin(kx)\left(a^\dagger-a\right).
  \label{Hsingle}
\end{gather}
Here $a$ is the operator annihilating a cavity photon while $U_0$ describes the light shift introduced by the atom-field coupling.
The depth of the lattice potential is $V_0$ and $\Delta_c$ denotes the detuning of the cavity resonance from the lattice field.
We can now rewrite $H_0$ using the ladder operators $B$ and $B^\dagger$ to get
\begin{equation}
  \label{H_single}
  H=\Omega B^\dagger B+\left(\Delta_\mathrm{c}+U_0\right) a^\dagger a+i\eta\left(a^\dagger-a\right)\left(B+B ^\dagger\right),
\end{equation}
where $\hbar=1$, $\Omega=k\sqrt{2V_0/m}$, $\eta=\ell\sqrt{\hbar V_0U_0}$, and $\ell\ll 1$ is the Lamb-Dicke parameter.

When the cavity resonance is tuned to the blue side of the laser according to $\Delta_\mathrm{c}+U_0=\Omega$, each scattered photon picks up an extra energy $\hbar\Omega$ by which the atomic motion is cooled. For sufficiently large $\Omega$ (deep potential) the offresonant heating process indicated by the gray path in Fig.~\ref{fig:one} (a) can be neglected.
Omitting the corresponding terms we can rewrite the Hamilton operator \eqref{H_single} in the interaction picture as
\begin{equation}
  H=i\eta\left(a^\dagger B-a\,B^\dagger\right)\,.
\end{equation}

Notice that $H$ conserves the sum of photons and vibrational quanta and besides cooling also the reverse process where a photon is absorbed from the cavity is possible.
However, when a photon leaves the cavity, it takes away energy and entropy from the system which cannot be reabsorbed.
Hence the atom will eventually end up in the ground state with the mode in the vacuum state \cite{Beige}.

Let us now generalize the single-particle Hamiltonian \eqref{Hsingle} into a second quantized form for many atoms following Ref.~\cite{ChristophPRL}, where we will consider the lowest two levels of the lattice wells corresponding to the first two levels (called $b$ and $c$) in the Wannier expansion of the atomic field operator [see Fig.~\ref{fig:one} (b)].
We assume here that the light shift $U_0$ introduced by the atoms is small so that we can drop terms proportional to $U_0$ in the second term of Eq.~\eqref{H_single}.
Thus the interaction Hamiltonian reads
\begin{equation}
  \label{Hint}
  H_\mathrm{gain}=i\eta\left(a^\dagger b^\dagger c-a\,b\,c^\dagger\right).
\end{equation}
Here, $b$ and $c$ are the bosonic field operators for the atomic laser and the source mode, respectively. The term $a^\dagger b^\dagger c$ describes a transition of an atom from the source mode to the laser mode while a photon is created in the cavity. Note that Eq.~\eqref{Hint} is equivalent to nondegenerate parametric amplification, which has been extensively studied for squeezed light generation \cite{Walls}. For parameters yielding squeezing we thus expect nonclassical atom-photon correlations. Upconversion (atomic vibrational heating) only occurs when both modes are excited and light loss causes a net downward transfer of atoms. Here we omitted direct atom-atom interaction terms, as we are dealing here with an all-optical setup, where we can expect only small occupation numbers per site.

An important ingredient in the model is a proper and consistent description of pumping (atom injection) into the source mode $c$. The simplest mechanism is atomic tunneling between lattice sites outside and within the cavity mode. As in general the tunneling in the lower atom-laser band $b$ is much slower than in the source mode, it can be neglected so that only excited band atoms move.
This is a coherent process and a possible Hamilton operator has the form $i\Gamma^\dagger c-i\Gamma c^\dagger$ where the operator $\Gamma$ models all coupled sites of the lattice potential outside the cavity. When their number is large enough and they are sufficiently occupied, their quantum state can be approximated by a coherent state.
Tracing out the corresponding levels from the density matrix, we get an effective pump Hamiltonian of the form
\begin{equation}
  \label{Hpump}
  H_\mathrm{pump}=i\mu\left(c^\dagger-c\right).
\end{equation}
Within this approximation the number of atoms in the system (modes $b$ and $c$) is not conserved.
Nevertheless the total atom number is conserved so that the atomic pump phase enters as an artifact with no observable consequences.
Equation \eqref{Hpump} gives thus a good approximation to injection by tunneling from a coherent reservoir covering the essential physics in a self-consistent way.
Similar to injection atoms are coupled out from the lower band. We will model this simply by a decay rate (analogous to cavity loss) which describes the coupling to an empty mode continuum as it would be realized by selective RF out-coupling in an extra magnetic field gradient.

Figure \ref{fig:one} (b) sketches the atomic dynamics including pump, excitation transfer, and damping of the laser mode (dark arrows).
To be more realistic we enlarge this idealized dynamics by upper-state losses and incoherent atomic transitions where no photon is emitted into the resonator mode.
This models residual spontaneous emission and atomic background collisions heating the atomic motion.
Here, affected ground-state atoms are incoherently transferred to the source mode while upper-state atoms are just removed from the system as indicated by the gray arrows in Fig.~\ref{fig:one} (b).

In total the master equation reads
\begin{equation}
  \label{me}
  \dot{\rho}=-i\left[H_\mathrm{gain}+H_\mathrm{pump},\rho\,\right]+\mathcal{L}\rho,
\end{equation}
where the Liouville operator $\mathcal{L}\rho$ consists of the loss terms
\begin{gather}
  \mathcal{L}_k\rho=\kappa_k\left(\left[k,\rho\,k^\dagger\right] + \left[k\,\rho,k^\dagger\right]\right),\,k=a,b,c
\end{gather}
and the heating term of the atom laser mode
\begin{gather}
  \mathcal{L}_{bc}\rho=\kappa_{bc}\left(\left[b\,c^\dagger,\rho\,b^\dagger c\right] + \left[b\,c^\dagger\rho,b^\dagger c\right]\right).
\end{gather}
Following standard procedures this operator equation can be converted into a c-number Fokker-Planck equation for the generalized $P$ representation, which yields the following set of Langevin equations,
\begin{subequations}
  \label{leq1}
  \begin{gather}
    \dot{\alpha} = -\kappa_a\alpha+\eta\,\beta^\dagger\gamma+\xi_a\,,\\
    \dot{\beta} = -\left[\kappa_b+\kappa_{bc}\left(\gamma^\dagger\gamma+1\right)\right]\beta+\eta\,\alpha^\dagger\gamma+\xi_b\,,\\
    \dot{\gamma} = -\left(\kappa_c-\kappa_{bc}\beta^\dagger\beta\right)\gamma - \eta\,\alpha\beta+\mu+\xi_c\,,
  \end{gather}
\end{subequations}
where $\xi_a$, $\xi_b$, and $\xi_c$ are stochastic forces with zero mean.

\begin{figure}
  \includegraphics[width=8cm]{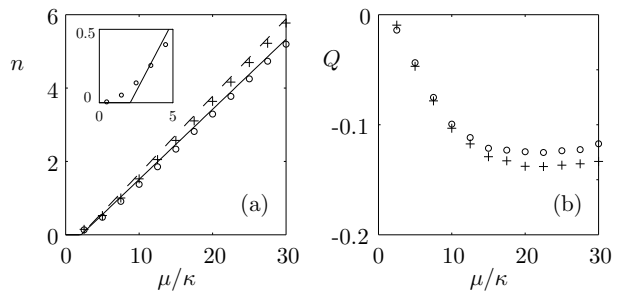}
  \caption{(a) Mean occupation number in the photon laser (+) and the atom laser ($\circ$) above threshold.
    The lines illustrate the semiclassical solution while the inset shows the behavior of the atom laser around threshold.
    (b) The corresponding Mandel-$Q$ parameter indicates sub-Poissonian statistics.}
  \label{fig:two}
\end{figure}

The semiclassical approximation to the dynamics amounts to ignoring the stochastic nature of Eqs.~\eqref{leq1},
so that for $\eta^2>\kappa_a\kappa_{bc}$ an oscillation threshold emerges at
\begin{equation}
  \mu_\mathrm{th}=\kappa_c\sqrt{\frac{\kappa_a\left(\kappa_b+\kappa_{bc}\right)}{\eta^2-\kappa_a\kappa_{bc}}}\,.
\end{equation}
Here the semiclassical steady-state solution for the field amplitudes shows an abrupt transition from zero to
\begin{subequations}
  \label{semi}
  \begin{gather}
    \alpha_0=\frac{\eta\sqrt{\kappa_c(\kappa_b+\kappa_{bc})}}{\eta^2-\kappa_a\kappa_{bc}}\sqrt{\epsilon-1}\,,\\
    \beta_0=\sqrt{\frac{\kappa_a\kappa_c}{\eta^2-\kappa_a\kappa_{bc}}}\sqrt{\epsilon-1}\,,
  \end{gather}
\end{subequations}
$\epsilon=\mu/\mu_\mathrm{th}$, while $\gamma_0=\mu/\kappa_c$ below the threshold and remains constant at $\mu_{\mathrm{th}}/\kappa_c$ above.
Let $n_a$ and $n_b$ denote the mean occupation numbers of the two laser modes.
When the atomic heating rate $\kappa_{bc}$ vanishes, we have
\begin{equation}
  n_k=\frac{\kappa_a\kappa_b\kappa_c}{\kappa_k\eta^2}\left(\epsilon-1\right),\,k=a,b,
\end{equation}
such that both output laser beams have the same mean intensity $I^\mathrm{out}=2\kappa_an_a=2\kappa_bn_b$.
As there is no other loss channel, this is even true for unequal decay rates $\kappa_a$ and $\kappa_b$.
Only the heating rate $\kappa_{bc}$ determines the ratio of the output-field mean intensities and it holds that $I_a^\mathrm{out}\geq I_b^\mathrm{out}$.
This is illustrated in Fig.~\ref{fig:two} (a) where $n_a$ and $n_b$ are plotted against the pumping strength $\mu$.
We chose the two output coupling rates to be equal (to $\kappa$), so that the output intensities are related to the mode occupation numbers by the same scaling parameter.
The other parameters are $(\kappa_{bc},\kappa_{c},\eta)=(0.1,10,5)\kappa$ and $\mu_\mathrm{th}\approx2.1\kappa$.
The markers correspond to the solution of the master equation \eqref{me} obtained by a Monte Carlo wave function simulation.
Each of the laser modes clearly exhibits an above-threshold population where the two solutions are in good agreement.

\begin{figure}[t]
  \includegraphics[width=8cm]{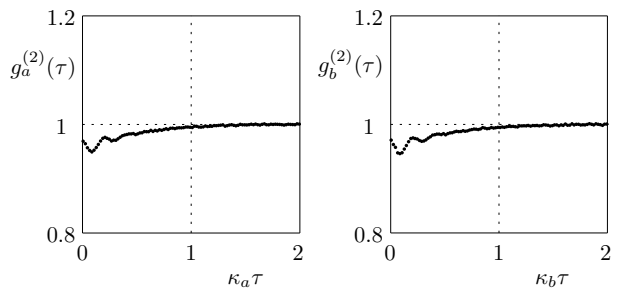}
  \caption{Second-order correlation function $g^{(2)}(\tau)$ of the photon laser (left) and the atom laser (right).}
  \label{fig:three}
\end{figure}

In addition to average intensities we will now consider intensity fluctuations.
For lasing the fluctuations of the intensity should be small compared to the mean.
To show this we have plotted the Mandel-$Q$ parameter, defined by $Q=\langle\Delta n^2\rangle/\langle n\rangle-1$, for the photon and atom statistics in Fig.~\ref{fig:two} (b).
While it should be close to zero for a laser, the negative $Q$ here even exhibits sub-Poissonian statistics.
In addition the requirement of second-order coherence can be recast in terms of Glauber's second-order time correlation function as
\begin{equation}
  |g^{(2)}(\tau)-1|\ll1.
\end{equation}
Figure \ref{fig:three} depicts the coherence function $g^{(2)}(\tau)$ for each laser mode where both lasers operate above threshold and $\kappa_a=\kappa_b$.

As mentioned our system more closely related to the non-degenerate parametric oscillator than a laser.
For photons the most striking consequence of parametric amplification is noise reduction of quadrature fluctuations and intensity correlations below the level of vacuum fluctuations \cite{Wu}.

\begin{figure}
  \includegraphics[width=8cm]{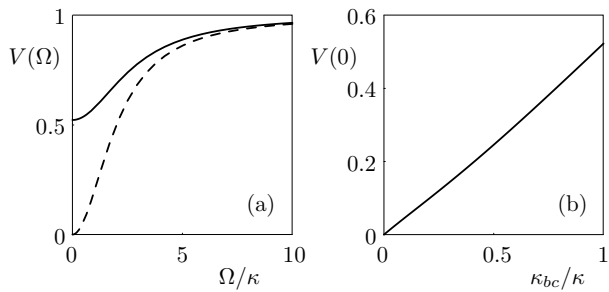}
  \caption{(a) Noise spectrum $V(\Omega)$ without heating (broken line) and with finite heating rate $\kappa_{bc}=\kappa$ (full line).
    (b) Zero-frequency component of the noise spectrum.}
  \label{fig:four}
\end{figure}

Since the ``twin'' signal and idler photons are created simultaneously by parametric down conversion, the fluctuations in the intensity difference are exactly zero in a system without dissipation.
When cavity decay (which is a random process affecting both modes independently) is included, this zero-noise feature can still be recovered in the output photocurrent difference, provided that all photons are detected and the detection time is sufficiently large.
Let us see what this means here. For $\kappa_{bc}=0$ analogous arguments will hold for our system, while the additional dissipation channel due to heating of ground-state atoms will break the perfect symmetry in the output fields and suppress noise reduction.

We now analyze the fluctuations in the output intensity difference $\Delta I=I_a^\mathrm{out}-I_b^\mathrm{out}$.
The noise spectrum, which is defined as
\begin{equation}
  V(\Omega)=\int\mathrm{d}\tau\,\mathrm{e}^{-i\Omega\tau}\left[\langle\Delta I(\tau)\Delta I(0)\rangle-\langle\Delta I\rangle^2\right],
\end{equation}
is plotted in Fig.~\ref{fig:four} (a).
As expected it shows perfect noise reduction at zero frequency for $\kappa_{bc}=0$ (broken line) while there is still squeezing of about 48\% below the shot noise level for $\kappa_{bc}=\kappa$ (full line).
The zero-frequency contribution $V(0)$ can be found in Fig.~\ref{fig:four} (b) as a function of $\kappa_{bc}$. Hence the systems realizes a quantum correlated atom-photon pair source.

In summary we showed that merging a photon laser and an atom laser in a cavity-confined optical lattice bears good prospects for a quantum source with genuine properties.
The stimulated emission driven Raman side-band cooling leads to lower temperatures than cavity cooling and suppresses the severe problems of spontaneous emission  and reabsorption. 
For sufficient pumping both quantum modes exhibit coherent oscillations with sub-Poissonian output statistics. While the laser outputs are second-order coherent, the output intensity difference shows noise compression very similar to parametric amplification and could be used to construct heralded coherent atom beams. 
We acknowledge funding by the Austrian Science Fund projects P17790 and S1512.

\end{document}